   \documentclass[aip,jcp,groupedaddress,a4paper,12pt,reprint]{revtex4-1} 
   \usepackage{amsmath}   %
   \usepackage{amsfonts}  
   \usepackage{amssymb}   %
   \usepackage{graphicx}  
   \usepackage[bookmarks=false]{hyperref} 
   \hypersetup{ 
    colorlinks=true,
    citecolor=blue,
    linkcolor=blue,
    urlcolor=blue}
   \usepackage{url} 
   \usepackage[all]{hypcap} 
   \usepackage{multirow} 

\begin{document}

\title{Ni-based nanoalloys: Towards thermally stable highly magnetic materials}

\author{Dennis Palagin}
\email{dennis.palagin@chem.ox.ac.uk}
\affiliation{Physical and Theoretical Chemistry Laboratory, Department of Chemistry, University of Oxford, South Parks Road, Oxford, OX1 3QZ, United Kingdom}

\author{Jonathan P. K. Doye}
\affiliation{Physical and Theoretical Chemistry Laboratory, Department of Chemistry, University of Oxford, South Parks Road, Oxford, OX1 3QZ, United Kingdom}

\date{29 September 2014}

\begin{abstract}
Molecular dynamics simulations and density functional theory calculations have been used to demonstrate the possibility of preserving high spin states of the magnetic cores within Ni-based core-shell bimetallic nanoalloys over a wide range of temperatures. We show that, unlike the case of Ni--Al clusters, Ni--Ag clusters preserve high spin states (up to 8\,$\mu_{\mathrm{B}}$ in case of Ni$_{13}$Ag$_{32}$ cluster) due to small hybridization between the electronic levels of two species. Intriguingly, such clusters are also able to maintain geometrical and electronic integrity of their cores at temperatures up to 1000\,K (e.g. for Ni$_{7}$Ag$_{27}$ cluster). Furthermore, we also show the possibility of creating ordered arrays of such magnetic clusters on a suitable support by soft-landing pre-formed clusters on the surface, without introducing much disturbance in  geometrical and electronic structure of the cluster. We illustrate this approach with the example of Ni$_{13}$Ag$_{38}$ clusters adsorbed on the Si(111)--(7$\times$7) surface, which, having two distinctive halves to the unit cell, acts as a selective template for cluster deposition.
\end{abstract}

\maketitle

\section{Introduction}

Metal clusters containing two different elements, known as bimetallic nanoalloys, attract much attention due to their specific properties and rich diversity of structures, which suggests possible applications in electronics, engineering, and catalysis.\cite{Ferrando_ChemRev_2008} Core-shell nanoalloys are especially interesting due to their chemical and physical properties that depend on the size and structure of the cluster as a whole, as well as the structure of the core and shell, and their interface in particular.\cite{Wei_Guo_review} Magnetic core-shell nanoclusters are of great importance because of the possibility of fine-tuning their properties to meet the needs of, for example, biotechnology applications.\cite{Wang_Zeng_review_2009, Jun_Choi_Cheon_review_2007} In particular, magnetic  nanoparticles have been suggested to be used in magnetic bio-separation,\cite{Lee_bioseparation} molecular imaging,\cite{Lee_bioimaging} or as bio-sensors.\cite{Grancharov_biosensors}

Ni-based clusters are very promising in this respect, as such compounds are known to exhibit interesting magnetic properties  in combination with other metals, such as, for example, Au,\cite{Chen_Shi_Du_2007} Pd,\cite{Guevara_2004} Fe,\cite{Robinson_Nguyen_2009} Cu and Co.\cite{Paola_Baletto_2013} Very attractive in this respect seem to be bimetallic clusters of Ni with Al and Ag.\cite{Paola_Baletto_2013, Deshpande_2010} Indeed, structure, as well as electronic, and magnetic properties of small Ni--Al\cite{Deshpande_2010, Li_Ren_2012, Wang_Huang_2009, Wen_Jiang_2010} and Ni--Ag\cite{Paola_Baletto_2013, Lee_Chen_2006, Baletto_Ferrando_2004, Harb_Simon_2010, Calvo_Wales_2013, Harb_Simon_2007, Springborg_2011, Rossi_Ferrando_2004, Rapallo_Johnston_2005} nanoalloys have recently been extensively investigated.

A crucial issue for the potential real life applications of such binary clusters is their stability (as well as the stability of their properties) towards changes in external conditions, especially temperature. For this reason, the study of thermal stability and phase transitions is important for design of novel nanoparticles with engineered properties.\cite{Haberland_2002, Aguado_2011} The thermal stability of individual Ni--Al\cite{Bailey_Johnston_2003, Chushak_2003, Jakse_Pasturel_2005, Wilson_Johnston_2006, Tang_Yang_2013} and Ni--Ag\cite{Calvo_Broyer_2008, Kuntova_Ferrando_2008, Mottet_Ferrando_2005, Oderji_Ferrando_2013} nanoalloy clusters has been theoretically investigated over the last decade.

Another important aspect of cluster studies in terms of practical applications is to move from isolated individual clusters towards the properties of such clusters in a non-trivial environment, i.e.\;under the influence of other bonding partners, which renders the clusters an integral part of new engineered materials. Experimentally this can be realised through deposition of pre-formed clusters on extended surfaces,\cite{Harbich_review} building of cluster-assembled materials,\cite{Claridge_review} encapsulating clusters into metal-organic frameworks (MOFs),\cite{Meilikhov_MOF} or even in more exotic ways, such as the recently reported deposition of core-shell clusters on the surface of silica microspheres.\cite{HHPark_2013}

Despite the wealth of experimental and theoretical studies outlined above, to the best of our knowledge no systematic study of the intriguing magnetic properties of intermediate-size Ni-based binary clusters and their thermal stability has been performed so far. Furthermore, neither the interplay between electronic structure and thermal properties of these nanoalloys, nor the applicability of such bimetallic clusters as building blocks to create engineered materials has been thoroughly investigated. Of course, such task ideally calls for multiscale modelling,\cite{Horstemeyer_review} both in terms of the timescales of appropriate computational methods, and the spatial description of a system. All these open questions motivate the present study on the geometrical and electronic structure, and thermal properties of intermediate-size Ni--Al and Ni--Ag binary nanoalloy clusters, which follows the following logic.

First, we perform global geometry optimization of a range of intermediate-size bimetallic clusters to establish their ground-state structures. For this we employ empirical potentials (EP) followed by density functional theory (DFT) calculations. We then run molecular dynamics (MD) simulations to find the melting temperature of each cluster and to generate a pool of geometries characteristic of a given cluster at a given temperature. For the obtained cluster configurations we analyse the electronic structure of the cluster configurations obtained both at the ground state and at higher temperatures at the DFT level of theory to elucidate the nature of chemical bonding within such core-shell clusters, and to see whether these nanoalloys exhibit magnetic properties, suitable for practical applications. Here we also analyse the difference in electronic interaction between the different shell elements (Al or Ag) and the potentially magnetic Ni core. Finally, we consider the feasibility of creating ordered arrays of such clusters on a suitable support by adsorbing them on the Si(111)--(7$\times$7) surface.

\section{Theory}

To make sure that geometries of all individual clusters under consideration indeed represent their ground-state structures, we relied on basin-hopping (BH) based global geometry optimization \cite{Wales_1997, Wales_2000} using an empirical potential, followed by the local reoptimization of the obtained structures with DFT. All sampling calculations have been carried out with the \texttt{GMIN} package\cite{GMIN_paper} using the Gupta potential,\cite{Gupta_original} which has been proved successful for describing both geometrical structure\cite{Johnston_review_2003, Hsu_2006} and thermal behaviour\cite{Hsu_2008} of metallic and bimetallic clusters of various compositions, including Ni--Al\cite{Bailey_Johnston_2003} and Ni--Ag.\cite{Kuntova_Ferrando_2008} 

The general form of the Gupta potential reads as:
\begin{equation}
U_{\mathrm{tot}} = A \sum\limits_{j \neq i}^{N} \upsilon \left(r_{ij}\right) - \xi \sum\limits_{i=1}^{N} \sqrt{\rho \left(r_{i}\right)}\; ,
\label{equation_1}
\end{equation}
where the potential $\upsilon \left(r_{ij}\right)$ is pairwise and defined as 
\begin{equation}
\upsilon\left(r_{ij}\right) = \exp \left[-p \left( \frac{r_{ij}}{r_{0}} - 1 \right) \right]\; ,
\label{equation_2}
\end{equation}
and the function $\rho\left(r_{i}\right)$ is given by:
\begin{equation}
\rho\left(r_{i}\right) = \sum\limits_{j \neq i}^{N} \exp \left[-2q \left( \frac{r_{ij}}{r_{0}} - 1 \right) \right]\; ,
\label{equation_3}
\end{equation}
where $A$, $\xi$, $p$, $r_{0}$ and $q$ are potential parameters. The values used in this study were taken from Ref. 
\citenum{Bailey_Johnston_2003} for Ni--Al and Ref. \citenum{Calvo_Wales_2013} for Ni--Ag, and are summarized in Table\,\ref{table_1}.

\begin{table}
\centering
\begin{tabular}{c c c c c c}

\hline \hline
      & $A$/eV &  $\xi$/eV & $p_{ij}$ & $q_{ij}$ & $r_{0}$/{\AA} \\
\hline 
\multicolumn{6}{c}{Ni--Al parameters\cite{Bailey_Johnston_2003}} \\
\hline
Ni--Ni & 0.0376 & 1.0700 & 16.999   & 1.1890  & 2.4911 \\

Ni--Al & 0.0563 & 1.2349 & 14.997   & 1.2823  & 2.5222 \\

Al--Al & 0.1221 & 1.3160 &  \multicolumn{1}{r}{8.612}   & 2.5160  & 2.8637 \\
\hline
\multicolumn{6}{c}{Ni--Ag parameters\cite{Calvo_Wales_2013}} \\
\hline
Ni--Ni & 0.0958 & 1.5624 & 11.340   & 2.270   & 2.4900 \\

Ni--Ag & 0.0096 & 1.3400 & 11.095   & 2.725   & 2.6900 \\

Ag--Ag & 0.1031 & 1.1895 & 10.850   & 3.180   & 2.8900 \\
\hline \hline

\end{tabular}
\caption{Gupta potential parameters for Ni--Al and Ni--Ag nanoalloys.}
\label{table_1}
\end{table}

The ground-state structures, as well as several low-lying higher energy local minimum configurations for each cluster have been reoptimized at the DFT level. Reoptimization of these structures, and subsequent electronic structure analysis was carried out with the plane-wave DFT package \texttt{CASTEP}.\cite{CASTEP_reference} Electronic exchange and correlation was treated within the generalized-gradient approximation functional due to Perdew, Burke and Ernzerhof (PBE).\cite{PBE_reference} For comparison single-point calculations at the optimized PBE geometries were also performed at the hybrid functional level using the PBE0 \cite{PBE0_reference} functional; we note that these calculations did not change any of the conclusions derived and presented below on the basis of the PBE data. The core electrons were described by using ultrasoft pseudopotentials, whereas the valence electrons were treated with a plane-wave basis set with a cut-off energy of 300\,eV. Local structure optimization is done using the Broyden-Fletcher-Goldfarb-Shanno method \cite{Nocedal_Wrigth_Num_Opt} relaxing all force components to smaller than $10^{-2}$\,eV/{\AA}.

All molecular dynamics (MD) simulations have been carried out using the \texttt{DL\_POLY} package.\cite{DLPOLY_reference} All cluster systems have been simulated for 500\,ns at temperature intervals of 50\,K, each time starting from the global minimum. Additionally, 50\,ns was used for equilibration. The Gupta potential\cite{Gupta_original} was used to model the interactions between atoms. The Velocity Verlet algorithm\cite{Velocity_Verlet} was used to integrate the equations of motion in the NVT ensemble. The length of time step was chosen as 1\,fs. The Nos{\'e}-Hoover thermostat\cite{Nose, Hoover} with relaxation time of 0.01\,ps was used to conserve the temperature. For an efficient sampling of phase space, we applied the replica exchange\cite{Swendsen_REX, Earl_Deem_parallel_tempering_review} algorithm. Swaps between adjacent temperatures were attempted every 10\,ns, with the exchange probability given by
\begin{equation}
P = \mathrm{min} \left\{ 1, \exp \left( -\frac{ \left( V_{j} - V_{i} \right) \left( T_{j} - T_{i} \right) }{T_{i}T_{j}k_{\mathrm{B}}} \right) \right\}\; ,
\label{equation_4}
\end{equation}
where $V_{i}$ and $V_{j}$, and $T_{i}$ and $T_{j}$ are potential energies and temperatures of two adjacent replicas, respectively, and $k_{\mathrm{B}}$ is the Boltzmann constant.

The final heat capacity curves were obtained by applying the weighted histogram analysis method (WHAM)\cite{WHAM_reference} to the replica exchange data as formulated in Ref.\citenum{Chodera_WHAM}. WHAM allows the computation of expectation values of an observable at any temperature of interest, if the average of the observable over all configurations with a given potential energy is known: 
\begin{equation}
\langle A \rangle _{\beta} = \frac{\int A \left( V \right) \Omega \left( V \right) e^{ - \beta V} d V}{\int \Omega \left( V \right) e^{ - \beta V} d V} \; ,
\label{equation_5}
\end{equation}
where $A\left(V\right)$ is the average of an observable $A$ over all configurations with potential energy $V$, $\Omega\left(V\right)$ is the potential energy density of states, and $\beta$ is the inverse temperature ($1/k_{\mathrm{B}}T$). 

The solid-liquid phase transition can be clearly identified by an abrupt peak in the heat capacity curve. In the canonical ensemble, the constant-volume heat capacity $C_{\mathrm{v}}$ is related to the potential energy fluctuations as follows, according to the fluctuation-dissipation theorem:
\begin{equation}
C_{\mathrm{v}}\left( T \right) = \frac{3}{2}Nk_{\mathrm{B}} + \frac{\langle V^{2} \rangle _{t} - \langle V \rangle^{2}_{t}}{k_{\mathrm{B}}T^{2}} \; ,
\label{equation_6}
\end{equation}
where $V$ is the potential energy, and $N$ is the number of atoms.

\section{Results}

\subsection{Ni--Al clusters}

\begin{figure}
\includegraphics[trim = 15mm 15mm 90mm 230mm,clip,width=\linewidth]{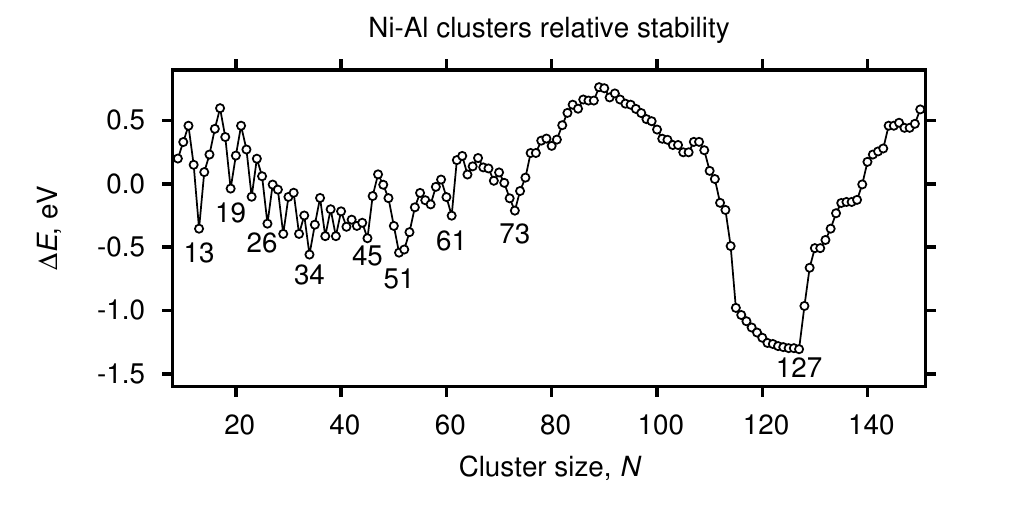}
\caption{Relative stability of Ni--Al clusters, computed according to the methodology introduced in Refs. \citenum{Rapallo_Johnston_2005} and \citenum{Doye_Meyer_2005}. See supplemental material\cite{supplement} for further details.}
\label{fig1}
\end{figure}

For the present study we have chosen nine Ni$_{n}$Al$_{m}$ clusters, corresponding to particularly stable configurations of the intermediate-size Ni--Al nanoalloys (cf.\,Fig.\,\ref{fig1}): Ni$_{1}$Al$_{12}$, Ni$_{2}$Al$_{17}$, Ni$_{4}$Al$_{22}$, Ni$_{7}$Al$_{27}$, Ni$_{12}$Al$_{33}$, Ni$_{12}$Al$_{39}$, Ni$_{13}$Al$_{48}$, Ni$_{19}$Al$_{54}$, and Ni$_{43}$Al$_{84}$. For the details on the procedure of assessing relative stabilities of the clusters please refer to supplemental material.\cite{supplement}

\begin{figure*}
\includegraphics[trim = 0mm 0mm 25mm 230mm,clip,width=0.9\textwidth]{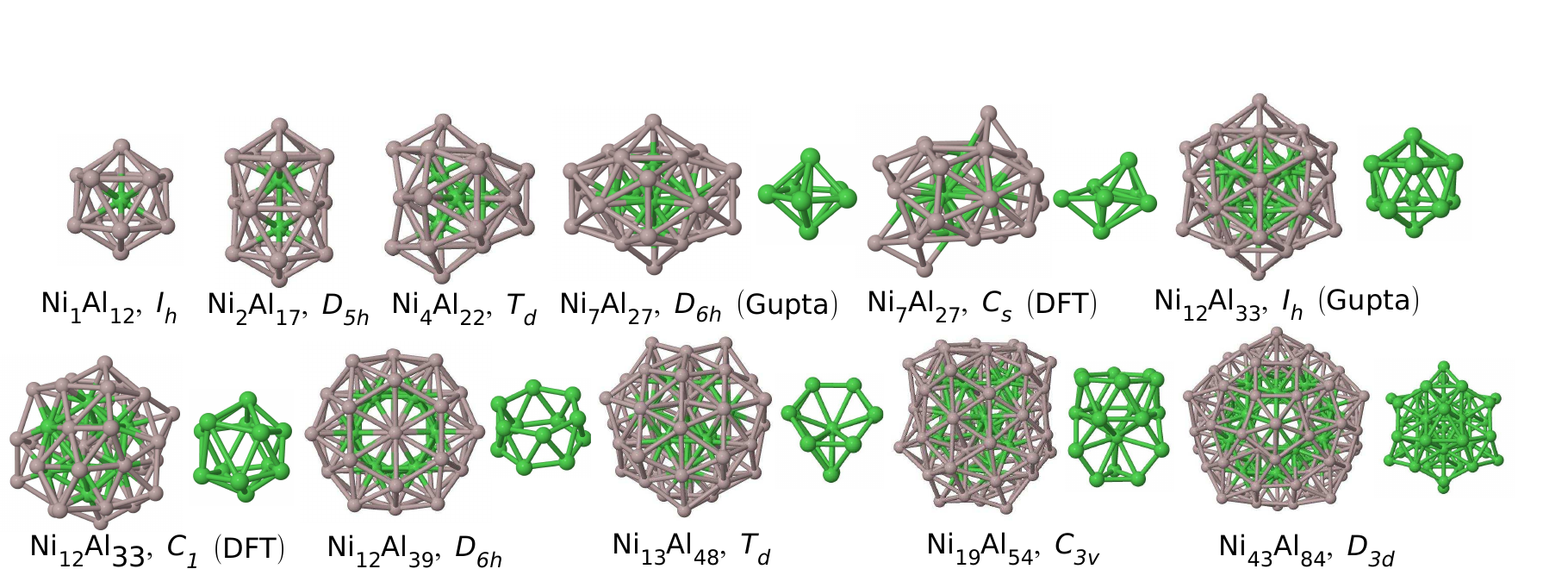}
\caption{Ground-state structures of Ni$_{n}$Al$_{m}$ clusters, identified by PBE DFT. Note the distorted structures of Ni$_{7}$Al$_{27}$ and Ni$_{12}$Al$_{33}$ predicted by DFT, as opposed to highly symmetrical geometries identified by Gupta potential.}
\label{fig2}
\end{figure*}

The identified ground-state structures (cf.\,Fig.\,\ref{fig2}) have also been reoptimized at the DFT level. These final geometries agree well with the ones obtained with the Gupta potential, except for two structures (Ni$_{7}$Al$_{27}$ and Ni$_{12}$Al$_{33}$), which are somewhat distorted in their ground states. Interestingly, the symmetry of the DFT optimized structure depends on the spin state. Thus, Ni$_{7}$Al$_{27}$ posesses a geometry of a very high  $D_{5h}$ point group symmetry when in a sextet spin state, while the doublet state has only $C_{s}$ symmetry. However, it is the doublet geometry that corresponds to the ground state, being 1.08\,eV more stable than its perfectly symmetrical sextet counterpart. A very similar picture is also observed in the case of Ni$_{12}$Al$_{33}$: here, the ground-state doublet structure symmetry is completely broken ($C_{1}$), while the 2.15\,eV less stable sextet is a perfectly symmetrical icosahedron ($I{_h}$). These results indicate strong interactions between the Ni and Al electronic levels, which prevent the system from preserving a high spin ferromagnetic alignment of the $d$-electrons in the Ni core, and thus lead to the distortion of the structure. The large impact of this effect on the geometry of the cluster is illustrated by the rather large energy differences between the structures. The origin of this coupling between geometry and electronic spin state is discussed in more detail below.

\begin{figure}
\includegraphics[trim = 15mm 15mm 60mm 190mm,clip,width=\linewidth]{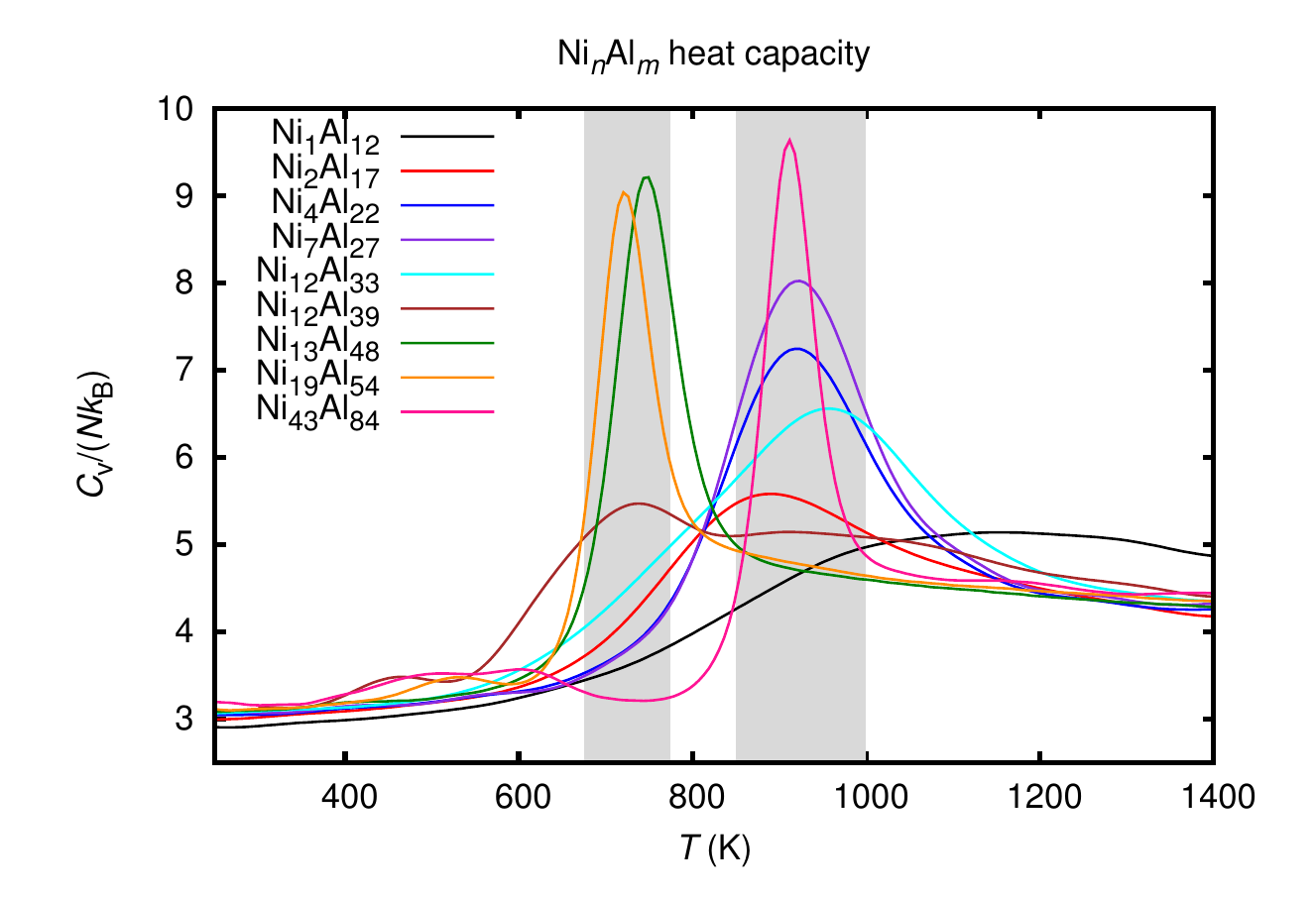}
\caption{Heat capacity curves for Ni$_{n}$Al$_{m}$ clusters. $N = n + m$.}
\label{fig3}
\end{figure}

The heat capacity curves for all nine considered clusters are summarized in Fig.\,\ref{fig3}. These heat capacity data were used to find the melting temperatures of the considered clusters.

Spin states of the Ni$_{n}$Al$_{m}$ clusters in their ground states and at their corresponding melting temperatures (main peaks) are summarized in Table\,\ref{table_2}. To find the spin states at high temperatures, the electronic structure of the most abundant configuration was computed at DFT level. Thus, these spin states can be referred to as typical spin states of the cluster at a given temperature.

\begin{table}
\centering
\begin{tabular}{c c c c}

\hline \hline
\multirow{2}{*}{Cluster} & Global minimum &  \multirow{2}{*}{$T_{\mathrm{melt}}$/K} & $T_{\mathrm{melt}}$ \\
                         & spin state     &                                          &    spin state       \\
\hline 
Ni$_{1}$Al$_{12}$  & triplet & 1150 & singlet \\
Ni$_{2}$Al$_{17}$  & doublet &  900 & doublet \\
Ni$_{4}$Al$_{22}$  & triplet &  920 & singlet \\
Ni$_{7}$Al$_{27}$  & doublet &  925 & doublet \\
Ni$_{12}$Al$_{33}$ & doublet &  950 & doublet \\
Ni$_{12}$Al$_{39}$ & sextet  &  740 & quartet \\
Ni$_{13}$Al$_{48}$ & septet  &  750 & singlet \\
Ni$_{19}$Al$_{54}$ & singlet &  740 & singlet \\
Ni$_{43}$Al$_{84}$ & singlet &  915 & singlet \\
\hline \hline

\end{tabular}
\caption{Melting temperatures and spin states of the Ni$_{n}$Al$_{m}$ clusters.}
\label{table_2}
\end{table}

\begin{figure}
\includegraphics[trim = 0mm 0mm 75mm 210mm,clip,width=\linewidth]{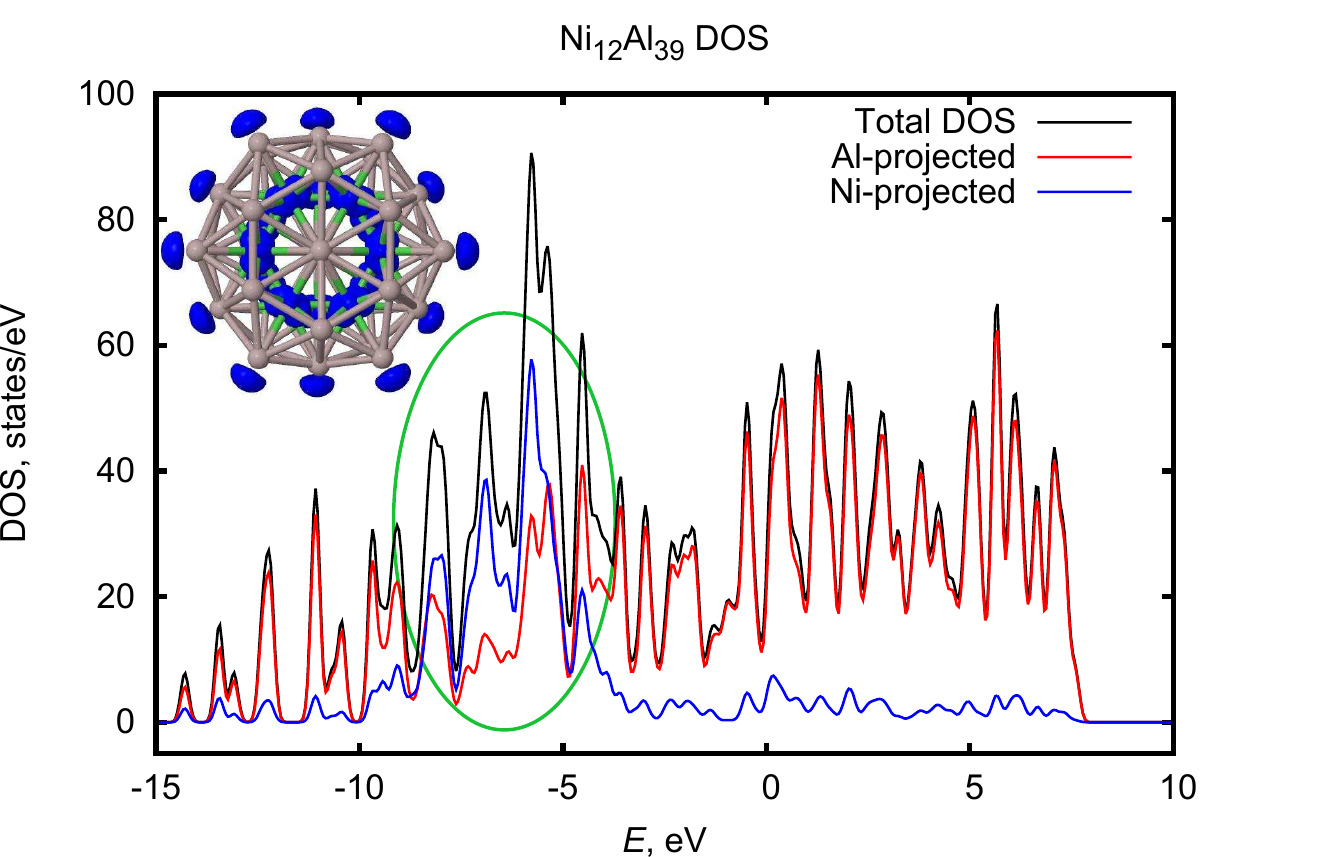}
\caption{Total density of states (black line) and DOS projected on Al shell (red line) and Ni core (blue line) atoms for the Ni$_{12}$Al$_{39}$ cluster. The highest occupied state lies at $-4.42$\,eV, where the vacuum
level is used as the zero reference. Frontier electronic levels are encircled with a green ring. The inset shows the spin density distribution within the structure.}
\label{fig4}
\end{figure}

As can be seen from the Table\,\ref{table_2}, high spin states are only observed for Ni$_{12}$Al$_{39}$ and Ni$_{13}$Al$_{48}$. At high temperatures, when distortion of the cluster structures is significant, the spin state is quenched in all cases. This can be rationalised in terms of hybridization between the electronic levels of Ni and Al. In Fig.\,\ref{fig4} we can see the density of states (DOS) diagram, where both levels of Ni and Al take part in forming the frontier electronic levels of Ni$_{12}$Al$_{39}$. As can be clearly seen from the 3D plots of the spin density distribution (see inset in Fig.\,\ref{fig4}), shell Al atoms also take part in accommodation of unpaired electrons, not only core Ni atoms.

For the case of Ni$_{13}$Al$_{48}$ both the DOS diagram and the spin density distribution look rather similar to the Ni$_{12}$Al$_{39}$ cluster (for the details on the electronic structure of the Ni$_{13}$Al$_{48}$ cluster please refer to supplemental material\cite{supplement}), with the hybridization between Ni and Al being even more pronounced. Interestingly, in both cases, the spin density is distributed over the Al atoms corresponding to the apices of the $T_{d}$ and $D_{6h}$ polyhedra of Ni$_{13}$Al$_{48}$ and Ni$_{12}$Al$_{39}$, respectively. Similar redistribution of electronic density might be responsible for geometrical distortion of the Ni$_{7}$Al$_{27}$ and Ni$_{12}$Al$_{33}$ clusters predicted by DFT as opposed to symmetrical Gupta structures, discussed above. 

Overall, strongly hybridized states render conservation of high magnetic moments of Ni core impossible, although clusters with unpaired electrons situated at the surface atoms (i.e. potentially unsaturated bonds), such as in case of Ni$_{12}$Al$_{39}$ and Ni$_{13}$Al$_{48}$, might be of interest for catalytic applications, as activation of clusters for catalysis is typically caused by coordinative (and, subsequently, electronic) unsaturation.\cite{Gates_review_2010}

\subsection{Ni--Ag clusters}

\begin{figure}
\includegraphics[trim = 15mm 15mm 90mm 230mm,clip,width=\linewidth]{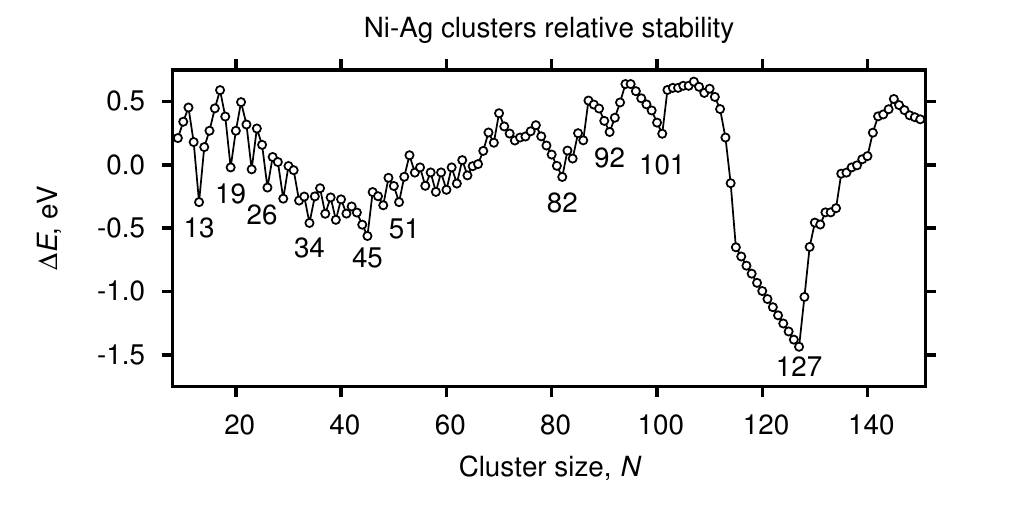}
\caption{Relative stability of Ni--Ag clusters, calculated similarly to Fig.\,\ref{fig1}.}
\label{fig5}
\end{figure}

\begin{figure*}
\includegraphics[trim = 0mm 0mm 25mm 180mm,clip,width=0.6\textwidth]{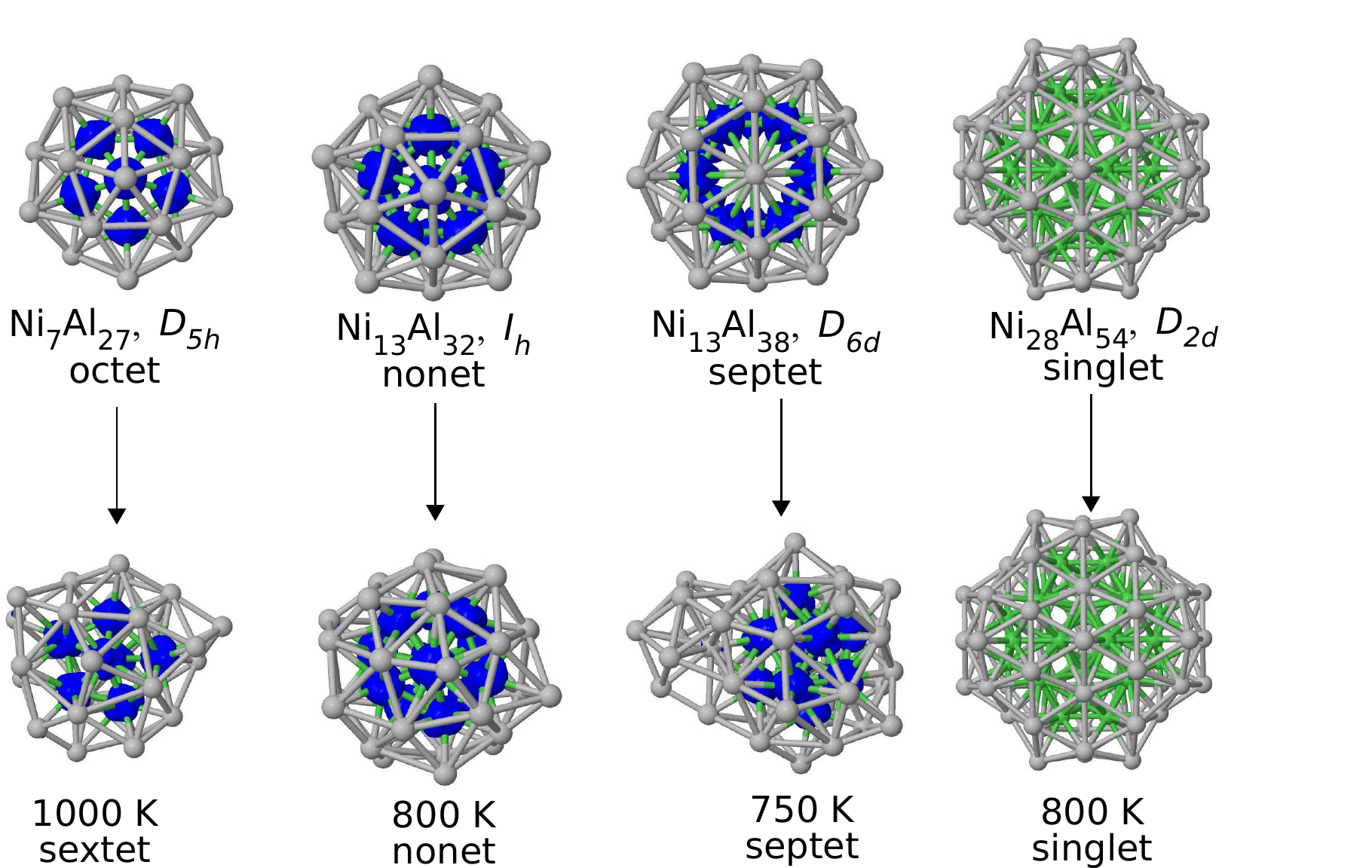}
\caption{Ground state structures and structures at $T_{\mathrm{melt}}$ of Ni$_{n}$Ag$_{m}$ clusters. Blue isosurfaces depict distribution of spin density.}
\label{fig6}
\end{figure*}

As we have seen with the example of Ni--Al binary nanoclusters, spin states of such systems are in most cases quenched due to rather strong hybridization between the electronic states of Ni and Al. Is it possible to overcome this problem by choosing another shell metal? Since small Ni--Ag nanoalloys have been reported to exhibit high spin states,\cite{Paola_Baletto_2013} we decided to investigate such clusters. Among the most stable Ni$_{n}$Ag$_{m}$ clusters (cf. Fig.\,\ref{fig5}) we have started from those having a potentially stable high spin core (Fig.\,\ref{fig6}).

Intriguingly, three of our selected clusters, Ni$_{7}$Ag$_{27}$, Ni$_{13}$Ag$_{32}$, and Ni$_{13}$Ag$_{38}$, exhibit very high spin states of octet, nonet and septet, respectively, with the spin density distributed exclusively among the Ni atoms. Even more striking is the fact that these high spin states are conserved at much higher temperatures, even up to 1000\,K for the case of Ni$_{7}$Ag$_{27}$. In the cases of Ni$_{7}$Ag$_{27}$ and Ni$_{13}$Ag$_{32}$, the configuration of the core remains virtually intact. For Ni$_{13}$Ag$_{38}$, there is a structural transition whereby the core adopts an icosahedral-like structure at high temperature, and the whole Ni$_{13}$Ag$_{38}$ cluster is like a capped version of the smaller Ni$_{13}$Ag$_{32}$ cluster. We will further discuss this interesting result later. The most massive nickel core in the largest selected cluster (Ni$_{28}$Ag$_{54}$) does not yield a high spin state.

\begin{figure}
\includegraphics[trim = 0mm 0mm 75mm 210mm,clip,width=\linewidth]{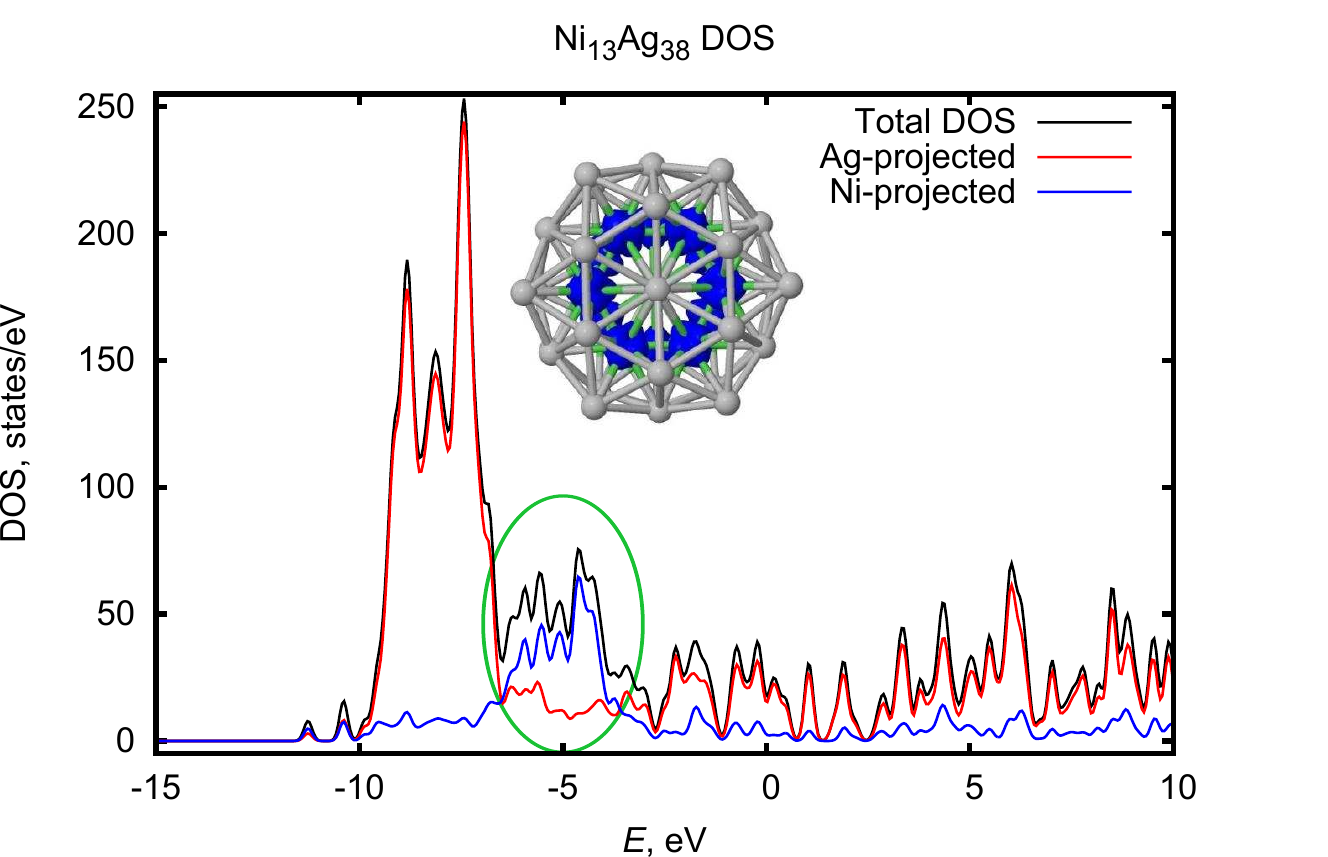}
\caption{Total DOS (black line) and DOS projected on Ag shell (red line) and Ni core (blue line) atoms for the Ni$_{13}$Ag$_{38}$ cluster. The highest occupied state lies at $-3.91$\,eV, where the vacuum level is used as the zero reference. Frontier electronic levels are encircled with a green ring. The inset shows the spin density distribution within the structure.}
\label{fig7}
\end{figure}

Analysis of the electronic structure (Fig.\,\ref{fig7}) shows a clear separation of electronic states of Ni and Ag atoms, e.g. in the case of Ni$_{13}$Ag$_{38}$. If we compare this diagram to the DOS of Ni$_{12}$Al$_{39}$ (Fig.\,\ref{fig4}), there is a striking difference in distribution of shell and core electrons. The circled region corresponds to the location of the cluster's frontier orbitals. Clearly, in the case of Ni$_{13}$Ag$_{38}$ cluster, the Ni and Ag states are rather well separated.

\begin{figure}
\includegraphics[trim = 0mm 0mm 80mm 205mm,clip,width=\linewidth]{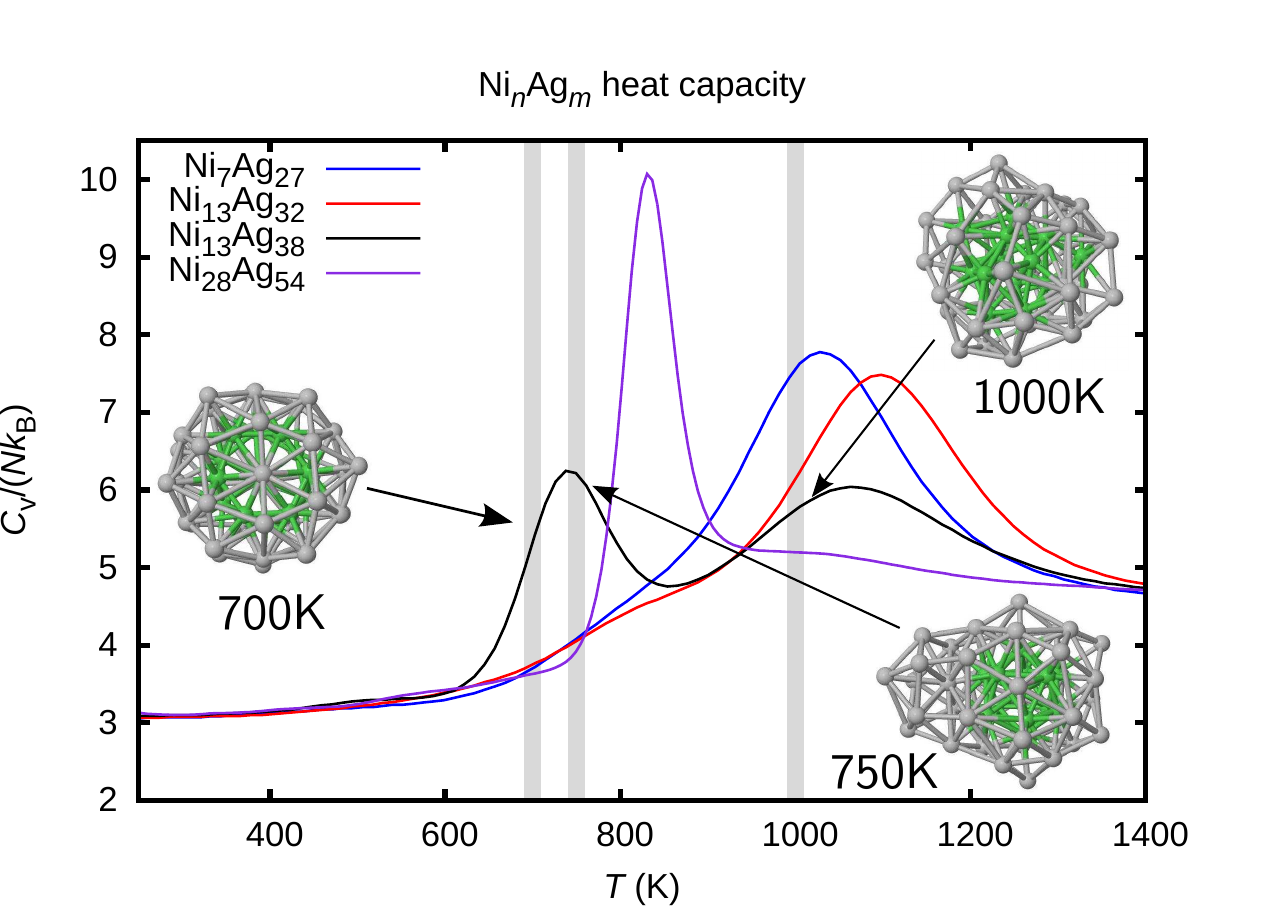}
\caption{Heat capacity curves for Ni--Ag clusters. Insets depict configurations of the Ni$_{13}$Ag$_{38}$ cluster at different temperatures. $N = n + m$.}
\label{fig8}
\end{figure}

The heat capacity curves for Ni--Ag clusters are rather similar to those of Ni--Al (cf.\,Fig.\,\ref{fig8} and Fig.\,\ref{fig3}). If we compare Ni$_{n}$Ag$_{m}$ clusters to their Al-based counterparts, we can see that the melting peaks are generally shifted towards higher temperatures in the case of Ag-based clusters. The most interesting $C_{\mathrm{v}}$ curve belongs to the Ni$_{13}$Ag$_{38}$ cluster, which undergoes a transition to a structure like that of Ni$_{13}$Ag$_{32}$ at high temperatures. There are two distinctive peaks at 740\,K and 1065\,K for the Ni$_{13}$Ag$_{38}$ cluster. At $\sim$750\,K reconfiguration of the core takes place from a hexagonal prism geometry to an icosahedron. Such a configuration is similar to the previously identified ground state of Ni$_{13}$Ag$_{32}$. For instance, one can clearly see the group of six ``extra'' Ag atoms that are expelled from the main cluster structure and onto the surface during the transition. At around 1000\,K the overall cage-like shape of the cluster is destroyed.  The position of the second peak (1065\,K) agrees well with the melting curve of Ni$_{13}$Ag$_{32}$. Interestingly, the core of the structure at 750\,K with a highly symmetric $I_{h}$ configuration still possesses a high spin septet state.

To rationalize such a reconfiguration of the Ni$_{13}$Ag$_{38}$ cage into a Ni$_{13}$Ag$_{32}$-like cage at higher temperatures, we computed the geometric mean vibrational frequencies of two configurations of the Ni$_{13}$Ag$_{38}$ cluster. Here, lower geometric mean vibrational frequency of the high temperature structure ($\bar{\nu}_{2}=196$\,cm$^{-1}$) compared to the global minimum structure ($\bar{\nu}_{1}=215$\,cm$^{-1}$) indicates the higher vibrational entropy, and hence increasing stability of the former structure as the temperature is increased. See supplemental material\cite{supplement} for a discussion of the full vibrational density of states for both structures.

The remarkable stability of the highly symmetrical magnetic (having nonet spin state in its ground state) Ni$_{13}$ cluster has been reported in the literature previously.\cite{Deshpande_2010, Vega_Alonso_1998, Grigoryan_2004} The important result here, however, is the fact that Ni$_{13}$ core surrounded by Ag shell atoms exhibits high magnetic moments in both hexagonal prism and icosahedron configurations, allowing the conservation of the septet spin state in Ni$_{13}$Ag$_{38}$ over a wide range of temperatures.

\subsection{Towards magnetic materials}

Thus, our results illustrate that, due to weak hybridization, Ni--Ag clusters of certain size exhibit high spin ground states, and are also able to preserve these spin states to higher temperatures. Is it thus plausible to suggest the potential applicability of such highly magnetic clusters as building blocks to create magnetic materials? The first obvious choice of how to realise such materials is to deposit pre-formed clusters on extended surfaces. Quite an extensive body of literature is devoted to adsorbed nanoparticles, most of which has been done in the context of catalysis.\cite{Gates_review_2003, Heiz_review_2004, Gates_review_2010} For instance, adsorption of bimetallic clusters on graphene,\cite{Yuan_Liu_Liu_2014} SiO$_{2}$,\cite{Wu_Jensen_2014} MgO,\cite{Ismail_Johnston_2013} TiO$_{2}$,\cite{Gulseren_2012} and Al$_{2}$O$_{3}$\cite{Nigam_2012} has been studied recently. The possibility of conserving and even increasing the high spin states upon adsorption was theoretically predicted for small Pd,\cite{Moseler_2002} Fe,\cite{Martinez_Illas_2009} and Rh\cite{Murugan_2006} clusters.

For the purpose of creating a structured array of adsorbed clusters, a suitable surface should: (i) be not very reactive towards our clusters (in order not to destroy their properties), and (ii) provide a certain template for adsorption, so that the clusters stay apart and do not aggregate. For example, the Si(111)--(7$\times$7) reconstructed surface\cite{Tong_Si7x7_1988, Joannopoulos_Si7x7_1992} might be suitable for this purpose, as most of the Si bonds are saturated there (but not all, thus allowing the possibility of moderately strong binding), and it has two distinctively different halves of the unit cell. If our clusters can predominantly be adsorbed on one of those, it will prevent them from aggregating.

\begin{figure}
\includegraphics[trim = 0mm 0mm 40mm 205mm,clip,width=\linewidth]{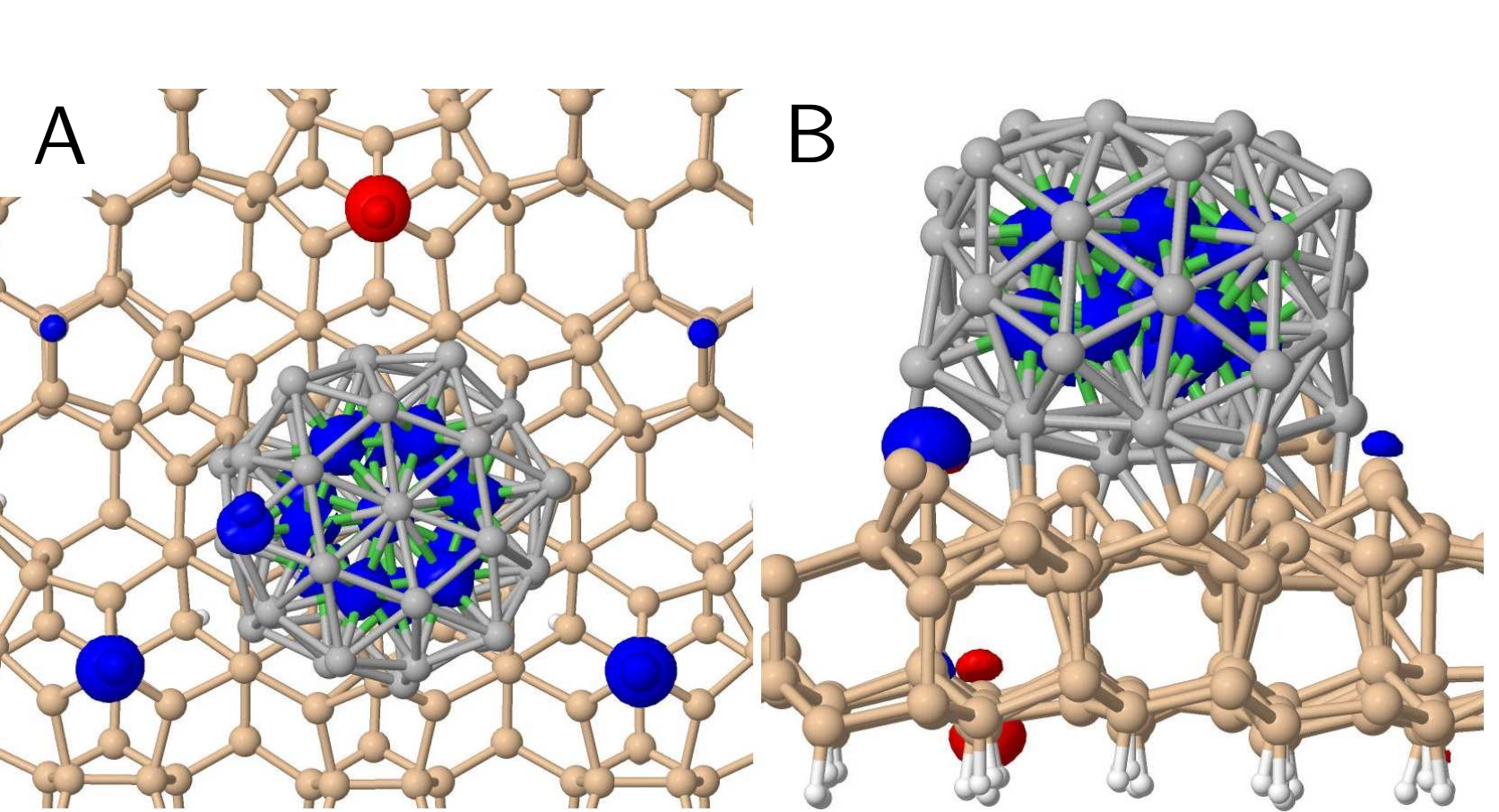}
\caption{Ni$_{13}$Ag$_{38}$ adsorbed on the unfaulted half of the unit cell (UHUC) of the Si(111)--(7$\times$7) surface: top view (panel A) and side view (panel B). Adsorption at the UHUC is 2.47\,eV more stable than at the FHUC. Isosurfaces depict spin density distribution. Blue and red isosurfaces depict alpha and beta spin channels, respectively.}
\label{fig9}
\end{figure}

Thus, we ran PBE DFT local geometry optimization of a cluster adsorbed on either the faulted half of the unit cell (FHUC) or the unfaulted half of the unit cell (UHUC) to check whether the cluster gets destroyed upon landing, and whether the two halves of the surface unit cell act differently. As potential adsorbates, we chose to investigate the Ni$_{13}$Ag$_{32}$ and Ni$_{13}$Ag$_{38}$ clusters. The Ni$_{13}$Ag$_{32}$ cluster turned out to be too stable individually, and it showed no tendency to adsorb. Ni$_{13}$Ag$_{38}$, on the other hand, allows for a certain rearrangement of electronic density between the cluster and the surface, which leads to adsorption energies of up to 13.38\,eV per cluster adsorbed on the unfaulted half of the Si(111)--(7$\times$7) unit cell (here positive adsorption energy indicates the energy gained upon adsorption, i.e.\;the total energy of the adsorbed aggregate is lower than the sum of the total energies of individual components). As can be seen from the Fig.\,\ref{fig9}, six unsaturated Si atoms (three adatoms and three rest atoms) interact with the adsorbed cluster. Taking into account the previously reported binding energy of $\sim 3.1$\,eV for a single Ag atom deposited on the Si(111)--(7$\times$7) surface,\cite{Zhang_Chen_Xiao_2005} the observed adsorption strength of our cluster can be characterized as moderate.

In the case of adsorption on either half of the unit cell, the adsorbed Ni$_{13}$Ag$_{38}$ cluster preserves its structural integrity, but there is a 2.47\,eV difference in binding energy between the two adsorption sites in favour of the UHUC. This result might seem surprising, as most indivudual metal atoms prefer the FHUC over the UHUC,\cite{Tanaka_Si7x7_2007} which was experimentally attributed to higher electron density (brighter scanning tunneling microscope (STM) images\cite{Joannopoulos_Si7x7_1994}) in the FHUC, and theoretically explained by the stability of metal atoms in the electrostatic wells produced by charge redistribution between Si adatoms and Si rest atoms.\cite{Kaxiras_Si7x7_1998} However, in a cluster as large as Ni$_{13}$Ag$_{38}$, surface shell atoms are already involved in sufficiently strong chemical bonding, which makes them in a sense ``less metallic''. In a similar way, for instance, Zn atoms with a fully filled $d$ level adsorb equally on the FHUC and UHUC by forming Zn$_{3}$ on the three center Si adatoms in each half unit cell.\cite{Tanaka_Si7x7_2007} In our case, adsorption on the FHUC leads to slightly larger distortions of the cluster, which turned out to be energetically less favourable. 

\begin{figure}
\includegraphics[trim = 10mm 0mm 60mm 175mm,clip,width=0.6\linewidth]{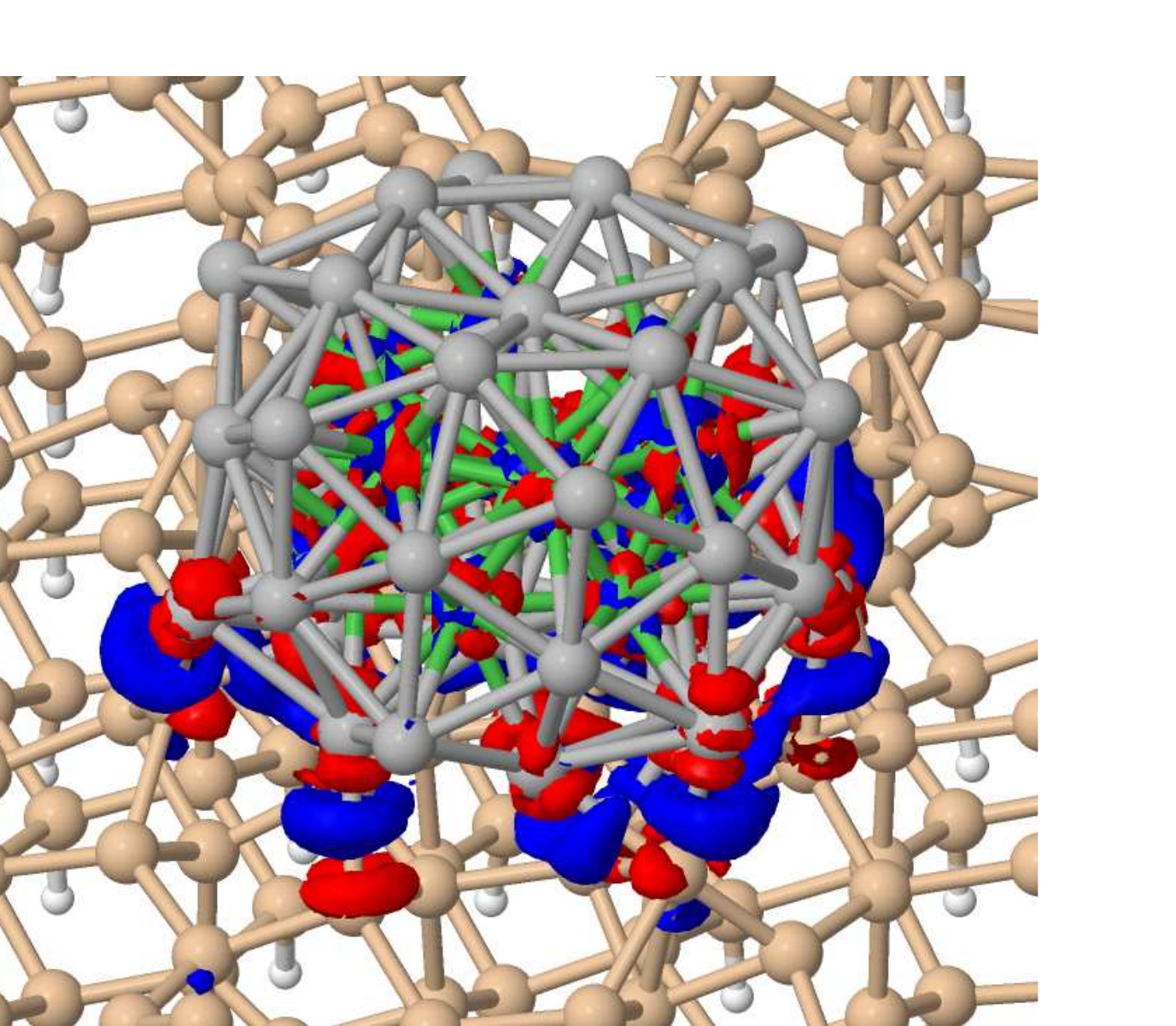}
\caption{Total electron density difference between the adsorbed cluster system and its components: $\Delta n = n\left(\mathrm{Ni}_{13}\mathrm{Ag}_{38} @\, \mathrm{surface}\right) - n\left(\mathrm{Ni}_{13}\mathrm{Ag}_{38}\right) - n\left(\mathrm{surface}\right)$. Blue isosurface depicts the gain of density, while red isosurface depicts the deficit.}
\label{fig10}
\end{figure}

If we take a look at the spin density distribution in the adsorbed cluster system (Fig.\,\ref{fig9}), the spin density isosurface plot reveals that most of the unpaired electrons are still situated at the Ni core of the cluster. The total spin density situated at core Ni atoms integrate up to 8\,$\mu_{\mathrm{B}}$, which is higher than the 6\,$\mu_{\mathrm{B}}$ in the individual Ni$_{13}$Ag$_{38}$ cluster. Such enhancement in spin states has been observed before in case of magnetic cluster adsorption,\cite{Moseler_2002, Martinez_Illas_2009} and has been explained by charge transfer from the cluster to the surface.\cite{Martinez_Illas_2009} A similar effect is also observed in our case. To illustrate this charge transfer, Fig.\,\ref{fig10} depicts the total electron density difference between the adsorbed cluster system and its components (i.e.\;bare surface and bare cluster): $\Delta n = n\left(\mathrm{Ni}_{13}\mathrm{Ag}_{38} @\, \mathrm{surface}\right) - n\left(\mathrm{Ni}_{13}\mathrm{Ag}_{38}\right) - n\left(\mathrm{surface}\right)$. Here, one can clearly see redistribution of the total electron density towards the surface (see blue areas in Fig.\,\ref{fig10}). 

Partial redistribution of the electron density indicates a certain level of interaction between the cluster and the surface, which helps to stabilize the adsorbed structure. In contrast, the more stable icosahedral Ni$_{13}$Ag$_{32}$ does not form stable adsorbed aggregates, due to insufficient interaction with the surface. We do not expect, however, that such inability of the symmetrical Ni$_{13}$Ag$_{32}$ cluster to adsorb can compromise the stability of adsorbed Ni$_{13}$Ag$_{38}$ clusters at higher temperatures, despite the reconfiguration of the Ni core for  the individual Ni$_{13}$Ag$_{38}$ cluster at 750\,K discussed above. First, the already formed bonds between the cluster and the surface would introduce an additional energy barier for reconfiguration. Secondly, the ``extra'' six silver atoms will have to either bind to the surface, or to adjacent shell atoms, either way contributing to the structure stability. However, the thermal behaviour of the assembled material is an important topic worthy of a separate investigation.

Intriguingly, UHUC adatoms take part in the spin density redistribution (Fig.\,\ref{fig9}). Taking into account the experimentally observed conductivity of the metallic dangling-bond states of adatoms on Si(111)--(7$\times$7) clean surface,\cite{Matsuda_2009} this might potentially mediate interaction between adsorbed clusters, which might be used for e.g. ferromagnetic coupling of their unpaired electrons. This question, however, goes beyond the scope of the current work, and we leave it here as an intriguing possibility.

Interestingly, the Ni$_{12}$Al$_{39}$ cluster also preferentially adsorbs on the UHUC of the Si(111)--(7$\times$7) surface. With the spin density of the individual cluster partially redistributed to the cluster's surface atoms, this leads us to suggest another potential use of such deposited clusters, namely as catalytic materials. Thus, the different hybridization behaviour of Al and Ag shell atoms with Ni core may be employed to create two different types of materials, i.e.\,either magnetic (weak hybridization, conserved spin states of the magnetic Ni core), or catalytic (strong hybridization, activated unsaturated shell Al atoms).

\section{Conclusions}

In summary, we have systematically studied the possibility of conserving the high spin states of the magnetic Ni core in Ni-based bimetallic nanoalloys with Al and Ag as shell metals. DFT-based electronic structure analysis revealed that Ni--Al bimetallic clusters are characterized by a strong hybridization between the electronic states of Ni and Al, which leads to quenching of the Ni core magnetic moment in most cases. In the case of Ni$_{7}$Al$_{27}$ and Ni$_{12}$Al$_{33}$ such hybridization leads to geometrical distortion of the high symmetry structures and complete quenching of the magnetic moment already in their ground states. This effect, apparently, is also responsible for the redistribution of the unpaired electrons, observed at the ground states of Ni$_{12}$Al$_{39}$ and Ni$_{13}$Al$_{48}$, over the surrounding shell Al atoms, which eventually leads to distortion of the clusters at higher temperatures.

Ni--Ag nanoalloys, however, exhibit a completely different behaviour. Ni$_{7}$Ag$_{27}$, Ni$_{13}$Ag$_{32}$, and Ni$_{13}$Ag$_{38}$ not only are magnetic in their ground state (octet, nonet, and septet, respectively), but also possess high spin states at higher temperatures: Ni$_{7}$Ag$_{27}$ retains a relatively high spin state of sextet at temperatures as high as 1000\,K, Ni$_{13}$Ag$_{32}$ is a nonet at 800\,K, and Ni$_{13}$Ag$_{38}$ is a septet at 750\,K. Ni$_{7}$Ag$_{27}$ and Ni$_{13}$Ag$_{32}$ clusters are able to mostly conserve their structural integrity at higher temperatures, whereas the Ni$_{13}$ core of Ni$_{13}$Ag$_{38}$ undergoes a transformation from a hexagonal prism to an icosahedral geometry at $T\sim750$\,K.

Furthermore, our calculations show the possibility of selective adsorption of such stable magnetic Ni--Ag clusters on the unfaulted half of the unit cell of the Si(111)--(7$\times$7) surface. For example, Ni$_{13}$Ag$_{38}$ prefers adsorption on the UHUC (2.47\,eV more stable than on the FHUC) while preserving integrity upon deposition. Even more interesting is the enhancement of the magnetic moment of the cluster up to 8\,$\mu_{\mathrm{B}}$ upon adsorption, compared to 6\,$\mu_{\mathrm{B}}$ in a gas phase. This offers the possibility of bringing the intriguing properties of such Ni--Ag bimetallic nanoalloys to real life applications via building novel materials with engineered properties.

Finally, different levels of interaction between Ni core and Al and Ag shell atoms leads to possible applications that are different in their chemical nature: Ni--Al as catalytically active materials, and Ni--Ag as magnetic ones. This illustrates the diversity and flexibility available due to the tunable nature of the properties of bimetallic nanoalloys, as opposed to the more ``predictable'' monometallic clusters, which suggests their applicability in the area of nanotechnology.

\section{Acknowledgements}

Funding within the EPSRC project \htmladdnormallink{EP/J011185/1}{http://gow.epsrc.ac.uk/NGBOViewGrant.aspx?GrantRef=EP/J011185/1} ``\htmladdnormallink{TOUCAN}{http://www.stchem.bham.ac.uk/~toucan/index.html}: TOwards an Understanding of CAtalysis on Nanoalloys'' is gratefully acknowledged. DFT calculations of the deposited clusters were carried out using the ARCHER UK National Supercomputing Service (\url{www.archer.ac.uk}). Less time intensive calculations were carried out at local compute facilities of Advanced Research Computing (\url{www.arc.ox.ac.uk}), University of Oxford.

\end{document}